\documentclass[aps,pra,twocolumn,superscriptaddress]{revtex4}
\usepackage{epsfig}
\usepackage{graphicx}
\usepackage{bm}

\begin{document}

\title{Creation of high-quality long-distance entanglement with flexible
resources}
\author{Bing He}
\email{bhe98@earthlink.net}
\affiliation{Department of Physics and Astronomy, Hunter College of the City University
of New York, 695 Park Avenue, New York, NY 10065}
\author{Yu-Hang Ren}
\affiliation{Department of Physics and Astronomy, Hunter College of the City University
of New York, 695 Park Avenue, New York, NY 10065}
\author{J\'{a}nos A. Bergou}
\affiliation{Department of Physics and Astronomy, Hunter College of the City University
of New York, 695 Park Avenue, New York, NY 10065}

\begin{abstract}
We present a quantum repeater protocol that generates the elementary
segments of entangled photons through the communication of qubus in coherent
states. The input photons at the repeater stations can be in arbitrary
states to save the local state preparation time for the operations. The
flexibility of the scheme accelerates the generation of the elementary segments 
(close to the exact Bell states) to a high rate for practical quantum communications. 
The entanglement connection to long distances is simplified and sped up, possibly realizing 
an entangled pair of high quality within the time in the order of that for classical communication 
between two far-away locations.
\end{abstract}

\maketitle

\section{Introduction} \label{section1}

The realization of quantum communications relies on setting up entanglement
of high fidelity between two far-away physical systems. In practice photons
are primarily used for the carrier of entanglement. If one tries to establish entanglement 
by sending photons directly to a remote place, however, the range of communication will be 
limited by their absorption losses in the transmission channel. A solution to the problem 
is quantum repeater \cite{repeater1,repeater2,childress}. There have been two
categories of physical approaches to realizing quantum repeater. One is DLCZ protocol \cite{DLCZ} 
and its developments \cite{simon,zhao,jiang,chen,sangouard, sangouard2}, which generate and connect
entangled pairs of atomic ensembles over short distances through the coupling of single photon and 
collective atomic excitation modes. The other is qubus or hybrid repeaters \cite{van-loock, ladd, 
munro, h-repeater} involving the operations on both qubits and continuous variable (CV) states. 
In the past years the developmental works have improved the
efficiency of the first type of quantum repeaters by several orders over the
original DLCZ protocol. A recent theoretical analysis \cite{sangouard2}, however, indicates that the minimum average time for distributing 
an entangled pair over $1200$ km by the quickest quantum repeater scheme of the DLCZ-type should be still more than a half minute. Moreover, the phase noise caused by birefringence and polarization mode dispersion on the traveling single photons could damage the quality of the generated pairs. 
With the qubus repeater protocols, on the other hand, the operation efficiency can be quickly improved at the 
cost of the fidelity of the generated pairs, and the irremovable decoherence effect on the CV state qubus in transmission channel is simply 
from photon absorption loss. 

In this work we present a new qubus repeater scheme combining some features 
of DLCZ-type repeaters. The resources (local input qubit state and memory space) required in the scheme
are flexible, and long distance entanglement with high quality can be quickly realized with such flexibility. 

\section{Purification of single photon sources} \label{section2}

\begin{figure}
\includegraphics[width=96truemm]{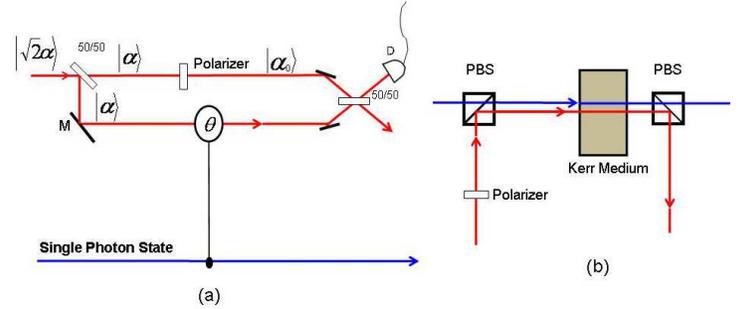}
\caption{Structure of QND module. (a) The interaction
of one of the coherent beams $|\alpha _{0}\rangle$ with a single
photon state in Kerr medium generates an extra phase $\theta $
on the coherent beam. The 50/50 beam splitter transforms two coherent states 
$|\alpha _{0}\rangle _{1}|\beta _{0}\rangle _{2}$, where $|\beta _{0}\rangle
_{2}=|\alpha _{0}\rangle _{2}$ or $|\alpha _{0}e^{i\theta
}\rangle _{2}$, to $|\frac{\alpha _{0}-\beta _{0}}{\sqrt{2}}\rangle _{1}|
\frac{\alpha _{0}+\beta _{0}}{\sqrt{2}}\rangle _{2}$. A response of the photodiode D 
indicates that the two beams $|\alpha _{0}\rangle _{1}$ and $|\beta _{0}\rangle _{2}$ are
different with $|\frac{\alpha _{0}-\beta _{0}}{\sqrt{2}}\rangle _{1}\neq
|0\rangle _{1}$. (b) Inside the Kerr medium, the single photon and the
coherent beam with the different polarizations are transmitted together.} 
\end{figure}

We start with the purification of the photon sources used in our scheme. The
output of a realistic single-photon source in a certain mode is approximated
by a mixture of single photon Fock state $|1\rangle$ and vacuum $|0\rangle$, 
$\rho=p_s|1\rangle\langle 1|+(1-p_s)|0\rangle\langle 0|$, where $p_s$ is the
efficiency of the source (see, e.g., \cite{source}). To sift the vacuum component out of the mixture, 
we apply a quantum non-demolition (QND) measurement module illustrated in Fig.
1. In the module, one of the laser beams in coherent state $|\alpha_0\rangle$
interacts with $\rho$ through a proper cross-Kerr nonlinearity, e.g., the electromagnetically 
induced transparency (EIT) medium, picking up a phase shift $\theta$ to $|\alpha_0 e^{i\theta}\rangle$ 
if the single photon is present. The two coherent states after the interaction are
compared with a 50/50 beam splitter and a photodiode. 
Any response of the photodiode indicates that the coherent states are different as $%
|\alpha_0\rangle$ and $|\alpha_0 e^{i\theta}\rangle$, projecting the mixture of photon and vacuum to a pure state of single photon. 
The success probability of the coherent states comparison is \cite{andersson}
\begin{eqnarray}
P_S=1-P_E=1-exp~(-\frac{1}{2}|\alpha_0-\alpha_0
e^{i\theta}|^2). 
\end{eqnarray}
If the coherent beam amplitude satisfies $|\alpha_0\theta| = 2\sqrt{5}$, e.g., the error probability $P_E$ will be as low as $e^{-10}$.
With the light-storage cross-phase modulation (XPM) technique, e.g., 
it is possible to realize a considerably large $\theta$ at a single photon level \cite{chen-yu}. 
In our scheme we only need a very small XPM phase shift $\theta \ll 1$, which is matched by the sufficiently intense coherent beams $|\alpha_0\rangle$, to lower the losses of the photonic modes in XPM process to a negligible level while achieving the close to unit $P_S$ in operation. 

Before the end of this section, we have a discussion on the photon detectors used in QND modules.
Here we only need threshold photon detector whose operation is described by the positive-operator-valued measure (POVM) elements $\Pi_0=\sum_{n=0}^{\infty}e^{-\lambda}(1-\eta_D)^n|n\rangle_1\langle n|$ and $\Pi_1=I-\Pi_0$, which respectively correspond to registering no photon and registering photon. The parameters $\eta_D$ and $\lambda$ are photon detection efficiency and average dark count during detecting photons, respectively. Since the mean dark count can be made small (a realistic detector could have $\lambda<10^{-6}$), the state of the photon source and the coherent beams in QND module will collapse to
\begin{widetext}
\begin{eqnarray}
\frac{\Pi_1^{\frac{1}{2}}(p_s|1\rangle\langle 1|\otimes|\beta_1\rangle_1\langle\beta_1|\otimes|\beta_2\rangle_2\langle\beta_2|+(1-p_s)|0\rangle\langle 0|\otimes|0\rangle_1\langle 0|\otimes|\sqrt{2}\alpha_0\rangle_2\langle \sqrt{2}\alpha_0|)\Pi_1^{\frac{1}{2}}}{Tr\{\Pi_1^{\frac{1}{2}}(p_s|1\rangle\langle 1|\otimes|\beta_1\rangle_1\langle\beta_1|\otimes|\beta_2\rangle_2\langle\beta_2|+(1-p_s)|0\rangle\langle 0|\otimes|0\rangle_1\langle 0|\otimes|\sqrt{2}\alpha_0\rangle_2\langle \sqrt{2}\alpha_0|)\Pi_1^{\frac{1}{2}}\}}
=|1\rangle\langle 1|\otimes \rho_{beam},
\label{detection}
\end{eqnarray}
\end{widetext}
the tensor product of a pure state single photon and that of the two beams $\rho_{beam}$, as the detector in QND module takes a response. 
Here $|\beta_1\rangle=|\frac{\alpha _{0}-\alpha _{0} e^{i\theta}}{\sqrt{2}}\rangle$ and $|\beta_2\rangle=|\frac{\alpha _{0}+\alpha _{0}e^{i\theta}}{\sqrt{2}}\rangle$. Given a very large amplitude $\alpha_0$ of the input coherent beams, the dominant part of the photon number Poisson distribution of $|\beta_1\rangle$ will be fairly away from the small photon numbers, and a realistic detector even with a low photon detection efficiency $\eta_D$ could obtain the output of Eq. (\ref{detection}) almost with certainty. Such detector can be simple photodiode.

\section{Pre-processing of Local Input Photons} \label{section3}

Next, we respectively process two purified single photons at two different
locations A and B with a linear optical circuit, as shown in Fig. 2 which outlines the setup to generate the elementary links. The purpose of the procedure is to transform an input photon pair in arbitrary rank-four mixed state, 
\begin{eqnarray}
\rho_{in}=\sum_{i=1}^4\sigma_i|\Lambda_i\rangle\langle\Lambda_i|,
\label{1}
\end{eqnarray}
to that with the linear combination of only two Bell states as the basis. The pure state components $|\Lambda_i\rangle$ as the eigenvectors of 
$\rho_{in}$ are the linear combinations of Bell states $|\Phi ^{\pm }\rangle=1/\sqrt{2}(|HH\rangle \pm
|VV\rangle )$ and $|\Psi ^{\pm }\rangle =1/\sqrt{2}(|HV\rangle \pm |VH\rangle)$, 
with $H$ and $V$ respectively representing the horizontal and the vertical
polarization, and $\sigma_i$ the eigenvalues of $\rho_{in}$. 
Its basis vectors $|\Phi^{\pm}\rangle$ are called even parity and $|\Psi^{\pm}\rangle$ odd parity, respectively. Then, with two polarization beam splitters (PBS), we convert any input into the ports $A1$ and $B1$ in Fig. 2 
to a which-path space. The polarization of the photon components on both path $1$ and $2$ can be
transformed to H with half wave plates (HWP or $\lambda/2$) for
the time being. The circuits A and B are constructed with 50/50 beam 
splitters, totally reflecting mirrors and the proper phase shifters, and any of the pure state component in a general
input $\rho_{in}$ will be mapped to the superposition of four bipartite
states over the pairs of output ports $\{K_A, K_B\}$, $\{R_A, R_B\}$, $\{K_A, R_B\}
$ and $\{R_A, K_B\}$ (see Appendix). Over port $K_A$ and $K_B$ ($R_A$
and $R_B$), the output is a linear combination of $|\Phi^-\rangle$
and $|\Psi^+\rangle$; over the other two pairs of ports $\{K_A, R_B\}$ and 
$\{R_A, K_B\}$, on the other hand, it is that of the other fixed set of
Bell states $\{|\Phi^+\rangle, |\Psi^-\rangle\}$. 
If we project out the photonic components over one of the four pairs of the output ports, the resulting 
state will be in a subspace with a linear combination of such two Bell states (one even
parity but the other odd parity) as the basis vector. Such projection can be done by two QND
modules (see Appendix). 

For example, by projecting out the components on $K_A$ and $K_B$ merged from the tracks $1^{\prime}$ and $%
2^{\prime}$ (at both locations) with HWP and PBS, we will realize the
following non-unitary transformations of the basis vectors: 
\begin{eqnarray}
&|\Phi^+\rangle\rightarrow 0,& ~~~~|\Phi^-\rangle\rightarrow \frac{1}{2}%
(|\Phi^-\rangle+i|\Psi^+\rangle),  \nonumber \\
&|\Psi^-\rangle\rightarrow 0,& ~~~~|\Psi^+\rangle\rightarrow \frac{1}{2}%
(|\Psi^+\rangle-i|\Phi^-\rangle).
\label{parity}
\end{eqnarray}
A simple input state $|HH\rangle=1/\sqrt{2}(|\Phi^+\rangle+|\Phi^-\rangle)$ 
will be correspondingly transformed to $(|H\rangle_A+i|V\rangle_A)(|H\rangle_B+i|V\rangle_B)$ 
with the common constant neglected.
With this example, we will demonstrate how to obtain an approximate Bell state
by separating the even and the odd parity sectors of the output.

\begin{widetext}

\begin{figure}
\includegraphics[width=140truemm]{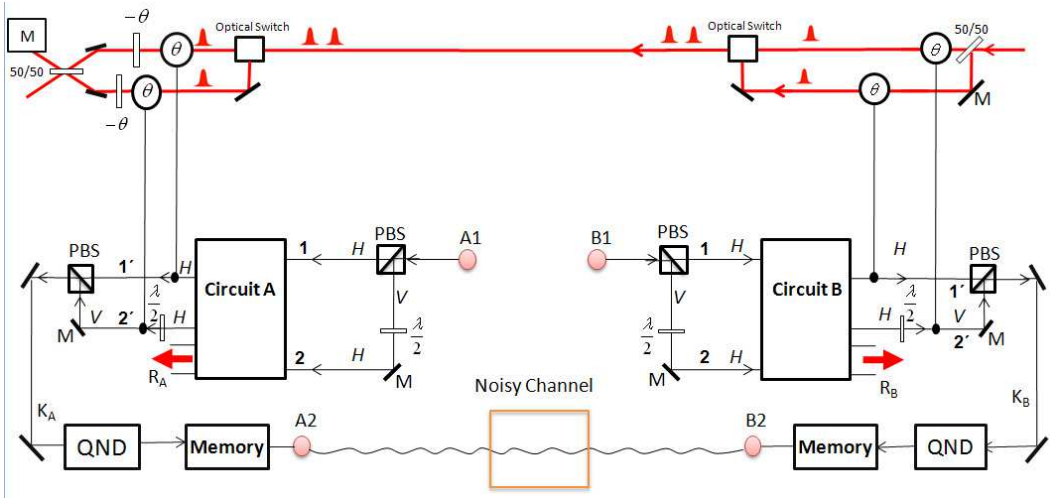}
\caption{Layout of setup for generating the elementary links. By two linear optical circuits A and B, each component 
$|\Lambda_i\rangle$ of a photon pair state in Eq. (\ref{1}), which is sent to A1-B1 terminals, is transformed to a superposition, $|\Sigma_i\rangle_{K_A,K_B}+|\Sigma_i\rangle_{K_A,R_B}+|\Sigma_i\rangle_{R_A,K_B}+|\Sigma_i\rangle_{R_A,R_B}$, 
over four pairs of output ports. Each of these states is the fixed linear combination of two Bell states (one even parity but the other odd). 
Through the Kerr media, two coherent pulse trains running between the output ports interact with the single photons at location A and B in the 
order specified here (the first coherent state is coupled to $H$ mode and the second to $V$ mode at location B, and they are coupled to the photons at location A in the opposite way), and separate the even and the odd parity components of the photon pairs. The detection of a local QND module heralds the single photon modes to be coupled to the traveling coherent beams and mapped to quantum memory. Upon receiving a successful result of the $M$ module and the measurement result of the QND module from location A, which are transmitted through a classical communication channel, the system operator at location B determines the generation and the type of an elementary link between A2-B2 terminals, together with the measurement of his/her local QND module.} 
\end{figure}

\end{widetext}

\section{generation of elementary links} \label{section4}

Meanwhile, as illustrated in Fig. 2, we interact two identical coherent beams $|\alpha\rangle$
with the photonic modes going to track $K_B$ through two XPM operations $U_{K_B, 1}=exp(i\chi t\hat{n}_H \hat{n}_{1})$ and $U_{K_B, 2}=exp(i\chi t\hat{n}_V \hat{n}_{2}$) ($\chi$ is the nonlinear intensity, $t$ the interaction time and $\hat{n}$ the number operator of the corresponding mode) by Kerr nonlinearities, evolving the state (projected out over tracks $K_A$ and $K_B$ here) from the input $|HH\rangle_{A,B} |\alpha\rangle_1|\alpha\rangle_2$ to $|\Psi_1\rangle=(|H\rangle_A+i|V\rangle_A)|\Psi'_1\rangle$, where
\begin{eqnarray}
|\Psi'_1\rangle &=& U_{K_B, 2}U_{K_B, 1}\{(|H\rangle_B+i|V\rangle_B)|\alpha\rangle_1|\alpha\rangle_2\} \nonumber\\ 
&=& U_{K_B, 2}(|H\rangle_B|\alpha e^{i\theta}\rangle_1|\alpha\rangle_2
+i|V\rangle_B |\alpha\rangle_1|\alpha\rangle_2)\nonumber\\ 
&=&|H\rangle_B|\alpha
e^{i\theta}\rangle_1|\alpha\rangle_2+i|V\rangle_B|\alpha\rangle_1|\alpha e^{i\theta}\rangle_2,
\label{a}
\end{eqnarray}
and $\theta=\chi t$.
If one also considers the losses of the photonic modes in the XPM processes, the above unitary operations will be replaced by non-unitary quantum operations leading to a mixed state similar to the form in Eq. (\ref{d}) below. In our case only a very small $\theta$ should be generated within a short interaction time $t$ of the coherent beams and the single photon, so the generated state has a close to unit fidelity with the pure state in Eq. (\ref{a}) \cite{bhe} and the XPM processes can be well approximated by the unitary operations. The coherent states in the above equation are transmitted through lossy
optical fiber to location A, while the single photon modes $\hat{a}_{H_B}$ and $\hat{a}_{V_B}$ are temporarily stored in quantum memory. Over a segment of fiber with the loss rate $\zeta$ of the coherent beams, their losses can be modeled by a beam splitter of transmission $\eta=e^{-\zeta L_0}$ for the distance $L_0$. Under such decoherence effect the initial state involving photon B and the coherent beams in Eq. (\ref{a}) will be decohered to \cite{state} 
\begin{eqnarray}
\rho_{B,1,2}=\frac{1+|\chi|^2}{2}|\Phi^1\rangle_{B,1,2}\langle\Phi^1|+\frac{%
1-|\chi|^2}{2}|\Phi^2\rangle_{B,1,2}\langle\Phi^2|,~  \label{d}
\end{eqnarray}
where 
\begin{eqnarray}
|\Phi^1\rangle_{B,1,2}&=&|H\rangle_B|\sqrt{\eta}\alpha e^{i\theta},\sqrt{\eta}\alpha
\rangle_{1,2} \nonumber \\
&+&i|V\rangle_B|\sqrt{\eta}\alpha,\sqrt{\eta}\alpha e^{i\theta} \rangle_{1,2},  \nonumber \\
|\Phi^2\rangle_{B,1,2}&=&|H\rangle_B|\sqrt{\eta}\alpha e^{i\theta},\sqrt{\eta}\alpha
\rangle_{1,2} \nonumber \\
&-&i|V\rangle_B|\sqrt{\eta}\alpha,\sqrt{\eta}\alpha e^{i\theta}\rangle_{1,2},  
\end{eqnarray}
and $\chi=\langle\sqrt{1-\eta}\alpha e^{i\theta}|\sqrt{1-\eta}\alpha\rangle$, after the beams are sent to location A. 
To eliminate the undesired component $|\Phi^2\rangle_{B,1,2}$ effectively, we could set the proper parameters such that $|\chi|^2\sim 1$. With the qubus coherent beams satisfying $|\alpha \theta|^2\sim 10^{-3}$, for instance, the fidelity $F$ with $|\Phi^1\rangle_{B,1,2}$ will be larger than $0.9995$.

After the coherent beams are transmitted to location A, we use the same Kerr
nonlinearities to interact the beams with the single photon modes there as shown in Fig. 2, realizing the state 
(the undesired contribution from $|\Phi^2\rangle_{B,1,2}$ component 
is eliminated effectively by the setting, $|\Psi_1(\eta)\rangle$ is the state with $\alpha$ replaced by $\sqrt{\eta}\alpha$ in Eq. (\ref{a}), and the indexes A, B are neglected too)
\begin{eqnarray} 
|\Psi_2\rangle&=& U_{K_A, 2}U_{K_A, 1}|\Psi_1(\eta)\rangle\nonumber\\
&=&U_{K_A, 2}(|HH\rangle|\sqrt{\eta}\alpha e^{i\theta}\rangle_1|\sqrt{\eta}\alpha \rangle_2\nonumber\\
&+&i|HV\rangle|\sqrt{\eta}\alpha \rangle_1|\sqrt{\eta}\alpha e^{i\theta}\rangle_2
+i|VH\rangle |\sqrt{\eta}\alpha e^{i2\theta}\rangle_1|\sqrt{\eta}\alpha \rangle_2\nonumber\\
&-&|VV\rangle|\sqrt{\eta}\alpha e^{i\theta}\rangle_1|\sqrt{\eta}\alpha e^{i\theta}\rangle_2)\nonumber\\
&=&|HH\rangle|\sqrt{\eta}\alpha e^{i\theta}\rangle_1|\sqrt{\eta}\alpha e^{i\theta}\rangle_2+i|HV\rangle|\sqrt{\eta}\alpha \rangle_1|\sqrt{\eta}\alpha e^{i2\theta}\rangle_2\nonumber\\
&+&i|VH\rangle |\sqrt{\eta}\alpha e^{i2\theta}\rangle_1|\sqrt{\eta}\alpha \rangle_2
-|VV\rangle|\sqrt{\eta}\alpha e^{i\theta}\rangle_1|\sqrt{\eta}\alpha e^{i\theta}\rangle_2.\nonumber\\
\end{eqnarray}
Then two phase shifters of $-\theta$ and one 50/50 beam splitter are applied to 
transform the state to  
\begin{eqnarray}  \label{k}
|\Psi_3\rangle&=&(|HH\rangle-|VV\rangle)|0\rangle_1|\sqrt{2\eta}%
~\alpha\rangle_2  \nonumber \\
&+&i|VH\rangle |i\sqrt{2\eta}~\alpha \sin\theta\rangle_1~|\sqrt{2\eta}%
~\alpha\cos\theta\rangle_2  \nonumber \\
&+&i|HV\rangle~|-i\sqrt{2\eta}~\alpha\sin\theta\rangle_1~|\sqrt{2\eta}%
~\alpha \cos\theta\rangle_2.~~~
\end{eqnarray}

Due to the setting of eliminating the decoherence effect, the detection rate of the first beam is very low if we measure 
it directly. For example, in the case of $|\alpha \theta|^2\sim10^{-3}$ and the elementary link distance $L_0=75$ km 
(the attenuation length of the coherent beams is assumed to be $25$ km), the intensity of $|\pm i\sqrt{2\eta}~\alpha \sin\theta\rangle_1$ will be as low as $10^{-5}$, which could be hardly detected in reality. Here we propose an indirect measurement approach by applying an extra QND module (denoted as $M$ in Fig. 2) in Fig. 1. In module $M$ the weak output coherent beam interacts with one of the bright beams $|\gamma\rangle$, realizing the following state ($\beta=\sqrt{2\eta}~\alpha$): 
\begin{eqnarray}  \label{h}
&&|\Psi_4\rangle\sim|\Phi^{-}\rangle |0\rangle_1 |\beta\rangle_2
|\gamma\rangle_{3} |\gamma\rangle_{4}  \nonumber \\
&&+i|VH\rangle\left(|0\rangle_1|\gamma\rangle_{3}|\gamma\rangle_{4}+ i\beta
\sin\theta |1\rangle_1|\gamma
e^{i\theta}\rangle_{3}|\gamma\rangle_{4}\right) |\beta \cos\theta\rangle_2 
\nonumber \\
&&+ i|HV\rangle \left(|0\rangle_1|\gamma\rangle_{3}|\gamma\rangle_{4}-i\beta
\sin\theta |1\rangle_1|\gamma
e^{i\theta}\rangle_{3}|\gamma\rangle_{4}\right) |\beta \cos\theta\rangle_2. 
\nonumber \\
\end{eqnarray}
The dominant non-vacuum component of the weak beam is single photon which validates the above approximation. To consider all possible output states $|0\rangle$ and $|\gamma (e^{ik\theta}-1)/\sqrt{2}\rangle$ 
($k\geq 1$) generated by the vacuum and non-vacuum components of the weak beam, we can use a sufficiently large $|\gamma|$ so that the overlaps of their photon number Poisson distributions are negligible to number-resolving detection. In our setting only simple photodiode is necessary because of the negligible occurring of the states with $k> 1$.

By any response of the photodiode in module $M$, together with the detection results of two QND detection modules in Fig. 2, an approximate $|\Psi^{-}\rangle$ with the fidelity larger than $1-(1-\eta)|\alpha \theta|^2/2$ is therefore created between the ports $K_A$ and $K_B$. Counting the possibilities over four pairs of output ports, we obtain the following success probability of realizing an approximate 
Bell state ($\theta \ll 1$): 
\begin{eqnarray}
P_g=\frac{1}{2}(1-e^{-2\eta|\alpha\sin \theta|^2})= \frac{1}{2}(1-(2F-1)^{\frac{2\eta}{1-\eta}}). 
\label{p}
\end{eqnarray}
This is the total probability of the odd parity sector minus that of the vacuum component of the first coherent state in it, which contributes to no-response of the photodiode in $M$ module. The pre-factor $1/2$ is due to the equal proportions of even and odd parity sectors from Eq. (\ref{parity}) and the other similar relations.

\begin{figure}
\includegraphics[width=82truemm]{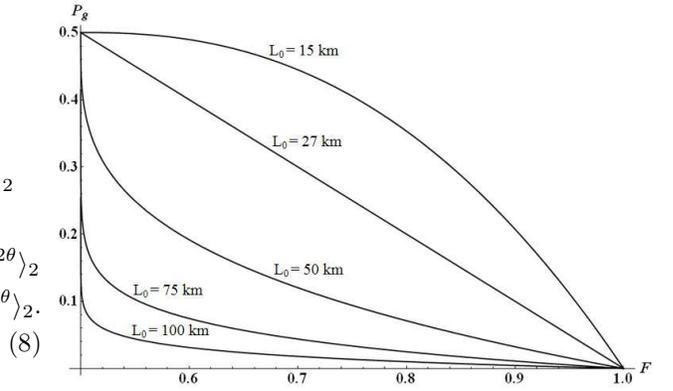}
\caption{The relation between $P_g$, the probability by a single try of entangling a photon pair, and $F$, the fidelity 
of the entangled pair. The distances between the repeater stations are chosen as $15$km, $27$km, $50$km, $75$km and $100$km.} 
\end{figure}

The efficiency of generating the elementary links of high fidelity is low as shown in Fig. 3. 
This can be overcome by the repeated operations with the frequency $f>c/(2L_0)$.
The input photons at location A and B can be in arbitrary states of Eq. (\ref{1}) even if $\sigma_i$ and $|\Lambda_i\rangle$ are unknown (see Appendix). It is therefore convenient to speed up the entangling operations with the continuous supply 
of single photons including the processed ones which can be recycled in case of a failure event (the losses of the photons in the XPM processes are neglected). Two pulsed laser beams with the repetition rate $f$, which are commonly generated in mode-locked system, realize an approximate Bell state within the average time $\tau/P_g+L_0/c$, the sum of the detection and the communication time (counted after the arrival of the first pulse), where $\tau=1/f$. Such repetitious operations should be matched by a quantum memory space $M_E=4[L_0f/c]$ ($[x]$ is the least integer equal to or larger than $x$) for the processed temporary single photon modes at location B, and a single photon source with the repetition rate $f$ is also necessary for starting such operations there. The memory modes of the successful events will be preserved while those of the failure are only stored for a time $2L_0/c$. 

To eliminate the difference of the phases of two qubus coherent pulse trains gained in travel, we transmit them through the same fiber within a very short time interval (Ref. \cite{DFS} presents an experimental study on the elimination of single photon phase noise in this way). Then the two pulse trains, which are controlled by optical switch (Pockels cell) at the starting and terminal point as in Fig. 2, act as the reference of each other to fulfill the purpose.

\section{Connection to long distances} \label{section5}

Given the high quality elementary links generated between the relay stations, we will connect them to doubled distances level by level through entanglement swapping. The local Bell state measurements here could be implemented by two-photon Hong-Ou-Mandel type interference \cite{zhao, jiang, chen, sangouard2}. The success probability $P_c=\eta^2_D \eta^2_M$ of a connection attempt, which is determined by those of retrieving two single photons from memory (with the efficiency $\eta_M$ for taking one photon out of a memory unit of two modes) and detecting both of them with the single photon detection efficiency $\eta_D$, is the same for all levels. On average, $[1/P_c]$ pairs at the $(n-1)$-th level are needed to realize an $n$-th level entangled pair with certainty and, between each two neighboring stations separated by the distance $L_0$, 
the total number of the required elementary links for such deterministic connection over the distance $L=2^n L_0$ should be therefore $[1/P_c]^n$. We adopt such strategy of connection: (1) generating so many elementary pairs between each two neighboring relay stations; (2) performing all local Bell state measurements iteratively at each connection level; (3) communicating the measurement results together to two link ends iteratively at every connection level. Summing up all these durations gives 
\begin{eqnarray}
T_{tot}&=&T_0[\frac{1}{P_c}]^n+\sum_{k=1}^{n}2^{k-1}\frac{L_0}{c}+\sum_{k=0}^{n-1}[\frac{1}{P_c}]^{n-k}\tau_0+\frac{L_0}{c} \nonumber\\
&=&T_0\left[\frac{1} {P_c}\right]^{log_2L/L_{0}}+\frac{\left[\frac{1}{P_c}\right]^{log_2L/L_{0}+1}-\left[\frac{1}{P_c}\right]}
{\left[\frac{1}{P_c}\right]-1}\tau_0+\frac{L}{c}\nonumber\\
\label{e}
\end{eqnarray}
as the average time for distributing an entangled pair over the distance $L$,  
where $T_0=L_0/c+\tau/P_g$ and $\tau_0\ll L_0/c$ is the time for a local Bell state measurement. 
The last term $L_0/c$ on the first line of the above equation is that for the first coherent pulse to arrive at location A. 
There are $5$ Bell state measurement results including failure. 
$T_{tot}$ is an upper bound time because elementary pairs generation can be performed simultaneously with connection. The final fidelity of a connected pair is $F'=F^{L/L_0}$ in terms of the elementary link fidelity $F$. With a close to unit initial fidelity ($F=1-x$, $x<10^{-3}$), $F'$ lowers slowly within the distance $L$ in the order of thousand kilometers.

\begin{figure}
\includegraphics[width=89truemm]{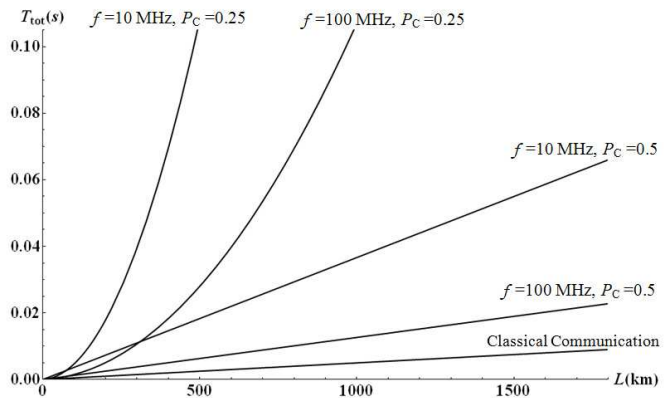}
\caption{The average times for distributing an entangled pair with some different coherent pulse repetition rates and connection probabilities. Here we assume $L_0=75$ km, $P_g=5\times 10^{-5}$, and $c=2\times 10^5$ km/s in optical fiber, and neglect the time for local Bell state measurements. By the connection strategy we present, $T_{tot}$ scales linearly if $P_c\geq 0.5$ and gets close to the two-way classical communication time with the also increased pulse train repetition rate $f$. The classical communication line is for comparison.} 
\end{figure}

From Fig. 4 we see that $T_{tot}$ can be reduced by adjusting $f$ and $P_c$ through two independent mechanisms. The two theoretical extreme values of $f$ respectively correspond to $T_0=2L_0/(cP_g)$ and $T_0=L_0/c+\tau_D/P_g$, where $\tau_D$ is the photodiode dead time of the QND modules in Fig. 2, and the necessary quantum memory spaces for the two cases are $M_E=2$ and $M_E=4[L_0/(c\tau_D)]$, respectively. We provide the following table of $T_{tot}$ and $M_E$ for distributing an entangled pair with 
the fidelity larger than $0.992$ over the distance of $1200$ km ($L_0=75$ km, $P_g=5\times 10^{-5}$, $F>0.9995$ as in the previous example, and $P_c=0.5$; the classical communication time over the distance is $6\times 10^{-3}$ s):
\begin{center}
\begin{tabular}{|c|c|c|c|c|c|c|c|}\hline
$f$ (Hz) & $1.33$ k & $40$ k & $1$ M & $10$ M & $100$ M\\ \hline
$T_{tot}$ (s) & $240$ &  $8$ & $3.3\times 10^{-1}$& $44\times 10^{-3}$& $15.2\times 10^{-3}$  \\ \hline
$M_E$ & $2$ &$60$ &$1.5\times 10^3$ &$1.5\times 10^4$ &$1.5\times 10^5$\\ \hline
\end{tabular}
\end{center}
The table demonstrates the trade-off between the required quantum memory coherence time and quantum memory space for the system. Even with the least temporary-mode memory space of $2$ per half station, such long-distance entanglement could be realized in $4$ minutes.

\section{Discussion} \label{section6}

We should mention another scenario that can be performed by the system: first generating the elementary pairs of the medium fidelity by a larger probability per try, and then performing the purification of the generated rank-two mixed states by local operations and classical communications \cite{pan} before entanglement swapping. Without using this entanglement purification step, the implementation of the scheme is much simplified, and the high efficiency can be achieved by the repetitious operations with the flexible input photon states. The cost to pay for this efficiency is a large memory space for the temporary modes. A number of recently developed multi-mode memories (see, e.g., \cite{simon, AFC, memory1, memory2}) could meet such requirement of the system in practical operations.

\begin{acknowledgments}
B. H. thanks C. F. Wildfeuer, J. P. Dowling for a material about Kerr nonlinearity, Y.-F. Chen, I. A. Yu for the discussions on experimental feasibility, and the comments from C. Simon.
This work is supported in part by the Petroleum Research Fund and PSC-CUNY award.
\end{acknowledgments}

\appendix

\section{Implementation of Non-unitary operations on Input Photon Pairs}
Since local operations and classical communication alone can not increase the entanglement of a bipartite system, we should perform non-local operation by two qubus coherent beams on the distant photons. Moreover, if a Bell state should be realized out of a photon pair in arbitrary rank-four mixed state of Eq. (\ref{1}), the application of the non-unitary transformations, those in Eq. (\ref{parity}) and the similar ones over the other pairs of output ports, on the input photon pair state will be essential. By these non-unitary transformations, all pure state components $|\Lambda_i\rangle$ of an input photon pair state are probabilistically mapped to a one-dimensional subspace with a linear combination of two fixed Bell states of the different parities as the basis vector. Then, the traveling qubus beams will separate the different parity components as in Sec. (\ref{section4}) and realize a Bell state in an ideal situation. We here apply a general method to implement non-unitary transformation on single photon states by unitary operation in extended space and projection to subspace \cite{A1}. In what follows, we illustrate the design that realizes such non-unitary maps. 

In Fig. 2, after the state of an input photon pair is converted to a which-path space by two PBS, its Bell-state basis vectors will be transformed to the following:
\begin{eqnarray}
|\Phi^{+}\rangle &\rightarrow & \frac{1}{\sqrt{2}}(\hat{a}^{\dagger}_{1}\hat{b}^{\dagger}_{1}+%
\hat{a}^{\dagger}_{2}\hat{b}^{\dagger}_{2})|0\rangle\equiv |B^1\rangle_{12,12}, \nonumber\\
|\Phi^{-}\rangle &\rightarrow & \frac{1}{\sqrt{2}}(\hat{a}^{\dagger}_{1}\hat{b}^{\dagger}_{1}-%
\hat{a}^{\dagger}_{2}\hat{b}^{\dagger}_{2})|0\rangle\equiv |B^2\rangle_{12,12},\nonumber\\
|\Psi^{+}\rangle &\rightarrow & \frac{1}{\sqrt{2}}(\hat{a}^{\dagger}_{1}\hat{b}^{\dagger}_{2}+%
\hat{a}^{\dagger}_{2}\hat{b}^{\dagger}_{1})|0\rangle\equiv |B^3\rangle_{12,12}, \nonumber\\
|\Psi^{-}\rangle &\rightarrow & \frac{1}{\sqrt{2}}(\hat{a}^{\dagger}_{1}\hat{b}^{\dagger}_{2}-%
\hat{a}^{\dagger}_{2}\hat{b}^{\dagger}_{1})|0\rangle\equiv |B^4\rangle_{12,12},~~~~~~~
\label{path}
\end{eqnarray}
where $\hat{a}^{\dagger}_{i}$ and $\hat{b}^{\dagger}_{i}$ are the creation operators of the which-path photonic modes at location A and B, respectively. Any pure bi-photon state as the linear combination of $|\Phi^{\pm}\rangle$ and $|\Psi^{\pm}\rangle $ is correspondingly transformed to 
\begin{eqnarray}
|\Lambda\rangle_{12,12}=\sum_{\mu=1}^{4}c_{\mu}|B^{\mu}\rangle_{12,12},
\label{a1}
\end{eqnarray}
where $c_{\mu}$ are the linear combination coefficients.
We here use $|\Lambda\rangle_{ij\cdots m,kl\cdots n}$ to represent a bi-photon pure state over the paths numbered $i$, $j$, $\cdots$, $m$ at location A and the paths numbered $k$, $l$, $\cdots$, $n$ at location B. The indexes in the sets $\{i,j,\cdots,m\}$ and $\{k,l,\cdots,n\}$ are in the ascending order, and the numbers in two sets can be different.
 
\begin{figure}
\includegraphics[width=66truemm]{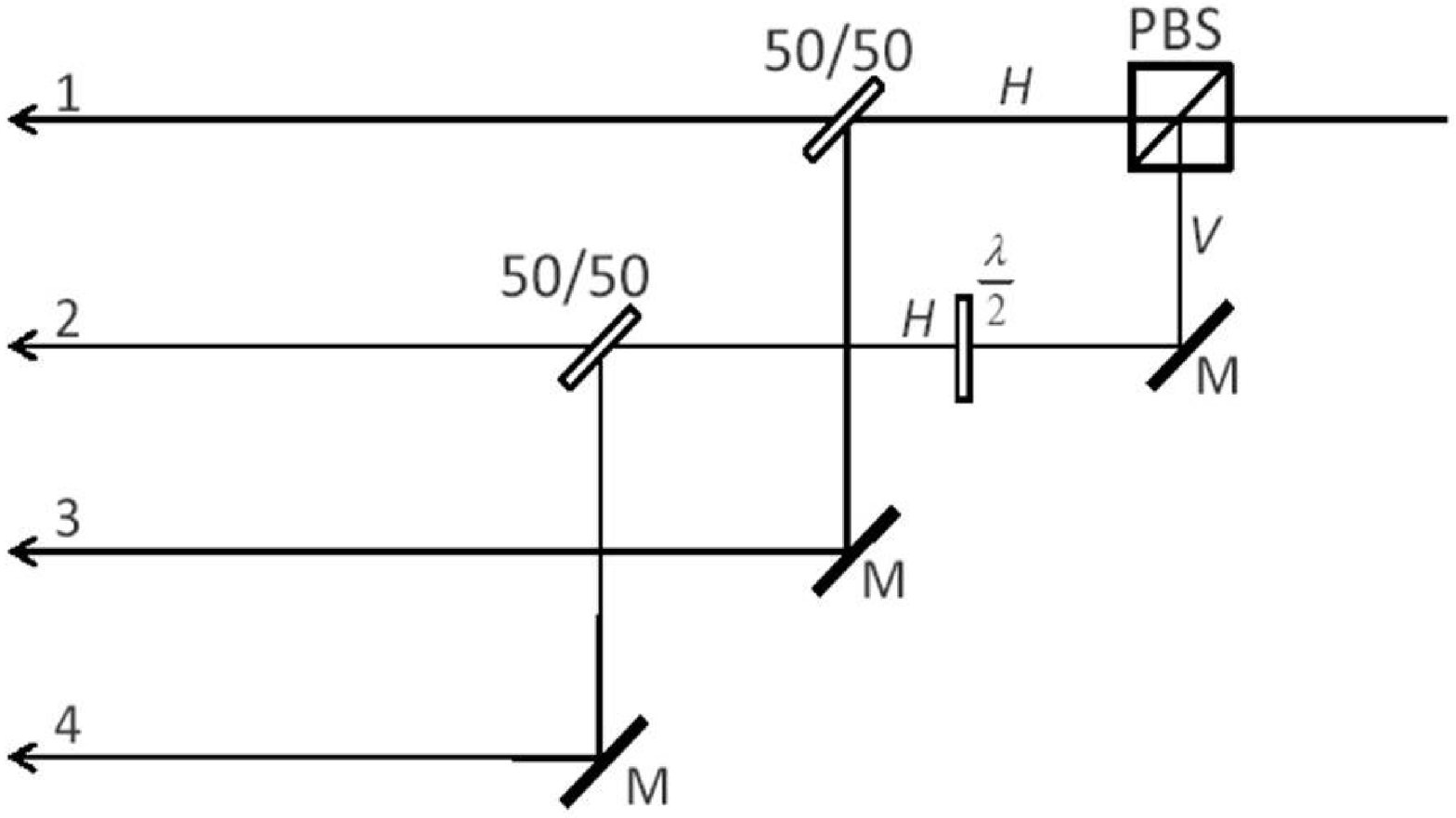}
\caption{The linear optical circuit that converts the input in polarization space to which-path space and implements $U_1$ of Eq. (\ref{a2}). 
The state of a single photon is converted to a which-path space by PBS. Then the mirrors M and the 50/50 beam-splitters transform the $2$-dimensional single photon state to a $4$-dimension one. It is a part of the integrated circuits A or B in Fig. 2.} 
\end{figure}

Each photon of a pair converted to the which-path space is sent into a linear optical circuit performing a unitary operation 
\begin{eqnarray}
U_1=\frac{1}{\sqrt{2}}\left(%
\begin{array}{cc}
I & I \\ 
-I & I%
\end{array}%
\right),
\label{a2}
\end{eqnarray}
in the extended $4\times 4$-dimensional space of the $2\times 2$-dimensional bi-photon state ($I$ in Eq. (\ref{a2}) means a $2\times 2$ identity matrix). Fig. 5 shows the local circuit performing this unitary operation. For technical simplicity, the unitary operations in the extended spaces of a bipartite state can be represented by a vector-operator duality notation as in \cite{A2} (the vector-operator duality description of all involved transformations of photon pair states here is given in \cite{reference}). We could also simply understand this transformation as one acting on a $4\times 4$ bi-photon state
\begin{eqnarray}
|\Lambda\rangle_{1234,1234}&=& |\Lambda\rangle_{12,12}+|null\rangle_{12,34}
+|null\rangle_{34,12}\nonumber\\
&+&|null\rangle_{34,34}
\end{eqnarray}
with the zero coefficient components $|null\rangle_{ij,kl}=(\sum_{m=i,j}\sum_{n=k,l}0\hat{a}^{\dagger}_m\hat{b}^{\dagger}_n)|0\rangle$ in the extended dimensions. The operation $U_1\otimes U_1$ on such an extended state $|B^1\rangle_{1234,1234}$ gives 
\begin{eqnarray}
|\Omega^1\rangle_{1234,1234}&=&
\underbrace{\frac{1}{2\sqrt{2}}(\hat{a}^{\dagger}_{1}\hat{b}^{\dagger}_{1}+%
\hat{a}^{\dagger}_{2}\hat{b}^{\dagger}_{2})|0\rangle}\limits_{\frac{1}{2}|B^1\rangle_{12,12}}  
-\underbrace{\frac{1}{2\sqrt{2}}(\hat{a}^{\dagger}_{1}\hat{b}^{\dagger}_{3}+\hat{a}%
^{\dagger}_{2}\hat{b}^{\dagger}_{4})|0\rangle}\limits_{\frac{1}{2}|B^1\rangle_{12,34}} \nonumber\\
&-&\underbrace{\frac{1}{2\sqrt{2}}(\hat{a}^{\dagger}_{3}\hat{b}^{\dagger}_{1}+\hat{a}%
^{\dagger}_{4}\hat{b}^{\dagger}_{2})|0\rangle}\limits_{\frac{1}{2}|B^1\rangle_{34,12}} 
+\underbrace{\frac{1}{2\sqrt{2}}(\hat{a}^{\dagger}_{3}\hat{b}^{\dagger}_{3}+\hat{a}%
^{\dagger}_{4}\hat{b}^{\dagger}_{4})|0\rangle}\limits_{\frac{1}{2}|B^1\rangle_{34,34}}.  \nonumber\\
\end{eqnarray}
The results of the operation $U_1\otimes U_1$ on all $|B^{\mu}\rangle_{12,12}$ take the same form as the above. 
It is not necessary to process the local input photons at location A and B simultaneously, 
since $U_1\otimes U_1|\Lambda\rangle_{12,12}=(U_1\otimes I) (I\otimes U_1)|\Lambda\rangle_{12,12}$.

The second pair of unitary operations on the states spanned by $\{|\Omega^{\mu}\rangle_{1234,1234}\}$ are the identical $4\times 4$ unitary operators,
\begin{eqnarray}
U_2=\left(%
\begin{array}{cc}
0 & iV \\ 
I & 0%
\end{array}%
\right)=\left(%
\begin{array}{cc}
iV & 0 \\ 
0 & I%
\end{array}%
\right)\left(%
\begin{array}{cc}
0 & I \\ 
I & 0%
\end{array}%
\right),
\end{eqnarray}
with the submatrix  
\begin{eqnarray}
V=\left(%
\begin{array}{cc}
0 & 1 \\ 
-1 & 0%
\end{array}%
\right),
\end{eqnarray}
to respectively act on the single photon modes at both locations. The consecutive operations of $U_1\otimes U_1$ and $U_2\otimes U_2$ on the Bell-state basis vectors $|B^{\mu}\rangle_{12,12}$ realize the states 
\begin{eqnarray}
|\Xi^{\mu}\rangle_{1234,1234}&=&\underbrace{\frac{1}{2}\{(\mp\frac{1}{\sqrt{2}}\hat{a}_1^{\dagger}\hat{b}_1^{\dagger}-\frac{1}{\sqrt{2}}\hat{a}_2^{\dagger}\hat{b}_2^{\dagger})|0\rangle\}}\limits_{\frac{1}{2}|A\rangle_{12,12}}\nonumber\\
&+&\underbrace{\frac{i}{2}\{(\mp\frac{1}{\sqrt{2}}\hat{a}_1^{\dagger}\hat{b}_4^{\dagger}+\frac{1}{\sqrt{2}}\hat{a}_2^{\dagger}\hat{b}_3^{\dagger})|0\rangle\}}\limits_{\frac{1}{2}|B\rangle_{12,34}}\nonumber\\
&+&\underbrace{\frac{i}{2}\{(\frac{1}{\sqrt{2}}\hat{a}_3^{\dagger}\hat{b}_2^{\dagger}\mp\frac{1}{\sqrt{2}}\hat{a}_4^{\dagger}\hat{b}_1^{\dagger})|0\rangle\}}\limits_{\frac{1}{2}|C\rangle_{34,12}}\nonumber\\
&+&\underbrace{\frac{1}{2}\{(\frac{1}{\sqrt{2}}\hat{a}_3^{\dagger}\hat{b}_3^{\dagger}\pm\frac{1}{\sqrt{2}}\hat{a}_4^{\dagger}\hat{b}_4^{\dagger})|0\rangle\}}\limits_{\frac{1}{2}|D\rangle_{34,34}}
\label{E}
\end{eqnarray}
for $\mu=1$ and $2$, and the states
\begin{eqnarray}
|\Xi^{\mu}\rangle_{1234,1234}&=&\underbrace{\frac{1}{2}\{(\pm\frac{1}{\sqrt{2}}\hat{a}_1^{\dagger}\hat{b}_2^{\dagger}+\frac{1}{\sqrt{2}}\hat{a}_2^{\dagger}\hat{b}_1^{\dagger})|0\rangle\}}\limits_{\frac{1}{2}|A\rangle_{12,12}}\nonumber\\
&+&\underbrace{\frac{i}{2}\{(\mp\frac{1}{\sqrt{2}}\hat{a}_1^{\dagger}\hat{b}_3^{\dagger}+\frac{1}{\sqrt{2}}\hat{a}_2^{\dagger}\hat{b}_4^{\dagger})|0\rangle\}}\limits_{\frac{1}{2}|B\rangle_{12,34}}\nonumber\\
&+&\underbrace{\frac{i}{2}\{(-\frac{1}{\sqrt{2}}\hat{a}_3^{\dagger}\hat{b}_1^{\dagger}\pm\frac{1}{\sqrt{2}}\hat{a}_4^{\dagger}\hat{b}_2^{\dagger})|0\rangle\}}\limits_{\frac{1}{2}|C\rangle_{34,12}}\nonumber\\
&+&\underbrace{\frac{1}{2}\{(\frac{1}{\sqrt{2}}\hat{a}_3^{\dagger}\hat{b}_4^{\dagger}\pm\frac{1}{\sqrt{2}}\hat{a}_4^{\dagger}\hat{b}_3^{\dagger})|0\rangle\}}\limits_{\frac{1}{2}|D\rangle_{34,34}}
\label{O}
\end{eqnarray}
for $\mu=3$ and $4$. $U_2$ is implemented by three totally reflecting mirrors and the phase shifters creating $\pm i$.

The third pair of local unitary operations is simply $U_3\otimes U_3=U_1^T\otimes U_1^T$ ($T$ represents the transpose). With the fourth pair of the local unitary operations, we extend the space of the bi-photon states further to $8\times 8$ dimension. The corresponding operator is an $8\times 8$ unitary matrix
\begin{eqnarray}
U_4=\left(%
\begin{array}{cc}
P_1 & P_2 \\ 
P_2 & -P_1
\end{array}%
\right)
\end{eqnarray}
constructed with the $4\times 4$ projection operators 
\begin{eqnarray}
P_1=\left(%
\begin{array}{cc}
0 & 0\\ 
0 & I
\end{array}
\right),~~~~ P_2=\left(%
\begin{array}{cc}
I & 0\\ 
0 & 0
\end{array}
\right).
\end{eqnarray}
Under all four successive local unitary operations, the Bell-state basis vectors will be transformed to
\begin{eqnarray}
&&\Pi_{i=1}^4 U_i\otimes U_i|B^{\mu}\rangle_{12,12}\nonumber\\
&=&\frac{1}{4}(|A\rangle_{34,34}+|B\rangle_{34,34}+|C\rangle_{34,34}+|D\rangle_{34,34})\nonumber\\
&+&\frac{1}{4}(|A\rangle_{34,56}-|B\rangle_{34,56}+|C\rangle_{34,56}-|D\rangle_{34,56})\nonumber\\
&+&\frac{1}{4}(|A\rangle_{56,34}+|B\rangle_{56,34}-|C\rangle_{56,34}-|D\rangle_{56,34})\nonumber\\
&+&\frac{1}{4}(|A\rangle_{56,56}-|B\rangle_{56,56}-|C\rangle_{56,56}+|D\rangle_{56,56}),~~~~~~
\label{basis}
\end{eqnarray}
with the states $|A\rangle$, $|B\rangle$, $|C\rangle$ and $|D\rangle$ respectively defined in Eqs. (\ref{E}) and (\ref{O}) for the even and odd parity 
Bell-state basis vectors. In Fig. 2, the indexes $3$, $4$, $5$ and $6$ are changed to $1'$, $2'$, $\cdots$, and all circuits performing the four consecutive unitary transformations are represented by the integrated circuits A and B. The components of the circuits are just 50/50 beam splitters, totally reflecting mirrors and the appropriate phase shifters.

After the photonic modes are converted back to polarization space as in Fig. 2, we will obtain the following transformations of the Bell-state basis vectors by substituting $|A\rangle$, $|B\rangle$, $|C\rangle$ and $|D\rangle$ in Eqs. (\ref{E}) and (\ref{O}) into Eq. (\ref{basis}):
\begin{eqnarray}
|\Phi^{+}\rangle &\rightarrow &  \underbrace{\frac{1}{2}(-|\Phi^+\rangle_{K_A, R_B}+i|\Psi^-\rangle_{K_A, R_B})}\limits_{\frac{1}{\sqrt{2}}|V_3\rangle}\nonumber\\
&& +\underbrace{\frac{1}{2}(-|\Phi^+\rangle_{R_A, K_B}-i|\Psi^-\rangle_{R_A, K_B})}\limits_{\frac{1}{\sqrt{2}}|V_4\rangle}\nonumber\\
|\Phi^{-}\rangle &\rightarrow &~~ \underbrace{\frac{1}{2}(|\Phi^-\rangle_{K_A, K_B}+i|\Psi^+\rangle_{K_A, K_B})}\limits_{\frac{1}{\sqrt{2}}|V_1\rangle}\nonumber\\
&& + \underbrace{\frac{1}{2}(|\Phi^-\rangle_{R_A, R_B}-i|\Psi^+\rangle_{R_A, R_B})}\limits_{\frac{1}{\sqrt{2}}|V_2\rangle}\nonumber\\
|\Psi^{+}\rangle &\rightarrow & ~~\underbrace{\frac{1}{2}(|\Psi^+\rangle_{K_A, K_B}-i|\Phi^-\rangle_{K_A, K_B})}\limits_{-\frac{i}{\sqrt{2}}|V_1\rangle}\nonumber\\
&& +\underbrace{\frac{1}{2}(|\Psi^+\rangle_{R_A, R_B}+i|\Phi^-\rangle_{R_A, R_B})}\limits_{\frac{i}{\sqrt{2}}|V_2\rangle}\nonumber\\
|\Psi^{-}\rangle &\rightarrow &  \underbrace{\frac{1}{2}(-|\Psi^-\rangle_{K_A, R_B}-i|\Phi^+\rangle_{K_A, R_B})}\limits_{\frac{i}{\sqrt{2}}|V_3\rangle}\nonumber\\
&&+\underbrace{\frac{1}{2}(-|\Psi^-\rangle_{R_A, K_B}+i|\Phi^+\rangle_{R_A, K_B})}\limits_{-\frac{i}{\sqrt{2}}|V_4\rangle}.
\label{bell}
\end{eqnarray}

Let us look at the effect of the system in Fig. 2 on any one of the pure state components, $|\Lambda_i\rangle=c^i_1|\Phi^+\rangle+c^i_2|\Phi^-\rangle+c^i_3|\Psi^+\rangle+c^i_4|\Psi^-\rangle$, in Eq. (\ref{1}). 
Without loss of generality, we have one of the beams, $|\delta\rangle_{A,1}$ and $|\delta\rangle_{B,1}$, in the QND modules to couple respectively to 
$K_A$ and $K_B$ modes, and two traveling coherent beams denoted as $|\alpha\rangle_1$ and $|\alpha\rangle_2$ are also coupled to $K_A$ and $K_B$ modes during the entangling operation shown in Fig. 2. The QND modules here are a type of the slightly modified one from Fig. 1, with a coherent beam coupling to both $H$ and $V$ modes in turn. Without considering the losses of the qubus beams $|\alpha\rangle_1$ and $|\alpha\rangle_2$ in travel, the state from the input $|\Lambda_i\rangle$ (following Eq. (\ref{bell})) evolves under the interaction with these coherent beams as follows:
\begin{widetext}
\begin{eqnarray}
&&\{\underbrace{(\frac{c^i_2}{2}-i\frac{c^i_3}{2})(|\Phi^-\rangle_{K_A, K_B}+i|\Psi^+\rangle_{K_A, K_B})}\limits_{|\Sigma_i\rangle_{K_A,K_B}}
+\underbrace{(\frac{c^i_2}{2}+i\frac{c^i_3}{2})(|\Phi^-\rangle_{R_A, R_B}-i|\Psi^+\rangle_{R_A, R_B})}\limits_{|\Sigma_i\rangle_{R_A,R_B}}\nonumber\\
&-&\underbrace{(\frac{c^i_1}{2}+i\frac{c^i_4}{2}) (|\Phi^+\rangle_{K_A, R_B}-i|\Psi^-\rangle_{K_A, R_B})}\limits_{|\Sigma_i\rangle_{K_A,R_B}}-\underbrace{(\frac{c^i_1}{2}-i\frac{c^i_4}{2}) (|\Phi^+\rangle_{R_A, K_B}+i|\Psi^-\rangle_{R_A, K_B})}\limits_{|\Sigma_i\rangle_{R_A,K_B}}\}\nonumber\\
&\otimes & |\alpha\rangle_1|\alpha\rangle_2 |\delta\rangle_{A,1}|\delta\rangle_{A,2}|\delta\rangle_{B,1}|\delta\rangle_{B,2}\nonumber\\
&\rightarrow & \frac{1}{\sqrt{2}}(\frac{c^i_2}{2}-i\frac{c^i_3}{2})\{(|HH\rangle_{K_A,K_B}-|VV\rangle_{K_A,K_B})|0\rangle_1|\sqrt{2}~\alpha\rangle_2+i|VH\rangle_{K_A,K_B} |i\sqrt{2}~\alpha \sin\theta\rangle_1~|\sqrt{2}~\alpha\cos\theta\rangle_2\nonumber\\
&+&i|HV\rangle_{K_A,K_B}|-i\sqrt{2}~\alpha\sin\theta\rangle_1~|\sqrt{2}~\alpha \cos\theta\rangle_2 \}
|\frac{\delta e^{i\theta}-\delta}{\sqrt{2}}\rangle_{A,1}|\frac{\delta e^{i\theta}+\delta}{\sqrt{2}}\rangle_{A,2}
|\frac{\delta e^{i\theta}-\delta}{\sqrt{2}}\rangle_{B,1}|\frac{\delta e^{i\theta}+\delta}{\sqrt{2}}\rangle_{B,2}\nonumber\\
&+&\frac{1}{\sqrt{2}}(\frac{c^i_2}{2}+i\frac{c^i_3}{2})(|H\rangle_{R_A}-i|V\rangle_{R_A})(|H\rangle_{R_B}-i|V\rangle_{R_B})|0\rangle_1|\sqrt{2}\alpha e^{-i\theta}\rangle_2 |0\rangle_{A,1}|\sqrt{2}\delta\rangle_{A,2}|0\rangle_{B,1}|\sqrt{2}\delta\rangle_{B,2}\nonumber\\
&-&\frac{1}{\sqrt{2}}(\frac{c^i_1}{2}+i\frac{c^i_4}{2}) \left(|H\rangle_{K_A}|\frac{\alpha e^{-i\theta}-\alpha}{\sqrt{2}}
\rangle_1|\frac{\alpha e^{-i\theta}+\alpha}{\sqrt{2}}\rangle_2 +i|V\rangle_{K_A}|\frac{\alpha -\alpha e^{-i\theta}}{\sqrt{2}}\rangle_1
|\frac{\alpha+\alpha e^{-i\theta}}{\sqrt{2}}\rangle_2 \right )|\frac{\delta e^{i\theta}-\delta}{\sqrt{2}}\rangle_{A,1}|\frac{\delta e^{i\theta}+\delta}{\sqrt{2}}\rangle_{A,2}\nonumber\\
&\otimes & (|H\rangle_{R_B}-i|V\rangle_{R_B}) |0\rangle_{B,1}|\sqrt{2}\delta\rangle_{B,2}\nonumber\\
&-&\frac{1}{\sqrt{2}}(\frac{c^i_1}{2}-i\frac{c^i_4}{2})(|H\rangle_{R_A}-i|V\rangle_{R_A})|0\rangle_{A,1}|\sqrt{2}\delta\rangle_{A,2}|\frac{\delta e^{i\theta}-\delta}{\sqrt{2}}\rangle_{B,1}|\frac{\delta e^{i\theta}+\delta}{\sqrt{2}}\rangle_{B,2}\nonumber\\
&\otimes &(|H\rangle_{K_B}|\frac{\alpha -\alpha e^{-i\theta}}{\sqrt{2}}
\rangle_1|\frac{\alpha+\alpha e^{-i\theta}}{\sqrt{2}}\rangle_2 +i|V\rangle_{K_B}|\frac{\alpha e^{-i\theta}-\alpha}{\sqrt{2}}\rangle_1
|\frac{\alpha e^{-i\theta}+\alpha}{\sqrt{2}}\rangle_2 ).
\label{output}
\end{eqnarray}
\end{widetext}
If we adopt a sufficiently large $|\delta|$ of the coherent beams, the overlap of $|0\rangle$ and $|(\delta e^{i\theta}-\delta)/\sqrt{2}\rangle$ will be small enough, and only one QND module at each location will be necessary. Then, after the interaction between $|\delta\rangle_{A,1}$, $|\delta\rangle_{B,1}$ and the local single photon modes, the responses of both QND modules at two locations project out the component $|\Sigma_i\rangle_{K_A,K_B}$ proportional to $|V_1\rangle$ defined in Eq. (\ref{bell}), as seen from Eq. (\ref{output}), and the other response and no-response patterns project out the similar components $|\Sigma_i\rangle_{R_A,R_B}$, $|\Sigma_i\rangle_{K_A,R_B}$, and $|\Sigma_i\rangle_{R_A,K_B}$, respectively. Also given the photon-number-resolving detection (Fock state projector) on the first traveling coherent beam finally output at location A and an enough large $|\alpha|$, $|\Phi^-\rangle$ or $|\Psi^{\pm}\rangle$ could be obtained between $K_A$ and $K_B$ depending on the projection on Fock states. From a general input state $\rho_{in}$ in Eq. (\ref{1}), the structures of the bi-photon states projected out by the QND modules are only relevant to the output ports; e.g., between $K_A$ and $K_B$, those from all pure state components $|\Lambda_i\rangle$ are proportional to 
$|V_1\rangle$ of Eq. (\ref{bell}). It implies that the finally output bi-photon state through the detections of the coherent beams will be a fixed Bell state independent of the input $\rho_{in}$, which is not necessary to be known to operations.

Were there no loss of the traveling coherent beams, a Bell state would be certainly realized by each entangling attempt following the operation order in Fig. 2 and collecting the possibilities from all four pairs of output ports. With which single photon modes the traveling beams should interact at one location is controlled by a detection result of the local QND module. Without such control with the classically feedfowarded measurement results, we can simply use two groups of coherent beams running between $\{K_A,K_B\}$ and $\{R_A,R_B\}$, respectively, to fulfill the same purpose at the price of some success probability per try. Due to the unavoidable losses of the traveling beams in optical fiber, we should adopt the setting discussed in Sec. \ref{section4} to maintain the high fidelity of the entangled bi-photon state. The relation between the efficiency and the fidelity in generating an entangled pair is then given in Fig. 3. In this case, the approximate Bell state generated between $K_A$ and $K_B$ will be $|\Psi^{-}\rangle$ because of the opposite phases of the coherent states $|i\sqrt{2\eta}~\alpha \sin\theta\rangle_1$ and $|-i\sqrt{2\eta}~\alpha \sin\theta\rangle_1$ in Eq. (\ref{k}), and only photodiode will be necessary in detection.

Finally, we provide a geometric interpretation for the involved operations here. Viewed from an extended space, the vectors $|\Lambda_i\rangle$ of Eq. (\ref{1}) are situated in a four-dimensional hyperplane spanned by $\{|\Phi^{\pm}\rangle, |\Psi^{\pm}\rangle\}$. The pairs of the local unitary transformations from $U_1\otimes U_1$ to $U_4\otimes U_4$ in the extended space rotate this hyperplane to a new position with the orthonormal vectors $\{|V_i\rangle\}$ ($i=1,\cdots,4$) defined in Eq. (\ref{bell}) being the basis of the rotated hyperplane. The circuits implementing these unitary operations $U_i$ are properly designed with a permutation symmetry in Eq. (\ref{path})---$|B^1\rangle$ and $|B^3\rangle$ are invariant while $|B^2\rangle$ and $|B^4\rangle$ change the sign under the permutation of the indexes $1$ and $2$, so that the new basis vectors $|V_i\rangle$ will be the linear combinations of two different parity Bell states. By all detection patterns of two QND modules, all $|\Lambda_i\rangle$ rotated to the hyperplane spanned by $\{|V_i\rangle\}$ are definitely projected onto one of the four lines in the direction of a $|V_i\rangle$. The $|V_i\rangle$ are separable because local operations can not increase the entanglement of a bipartite system. The parity gate with the operation of two qubus beams in Fig. 2 then separates the different parity components in a $|V_i\rangle$ to realize a Bell state in the ideal situation with no loss of the qubus beams in travel.

\end{document}